\documentclass[%
twocolumn,%
letterpaper,%
amsfonts,amssymb,amsmath,%
notitlepage%
]{revtex4-1}
\usepackage{graphicx}
\usepackage[breaklinks]{hyperref}
\usepackage[usenames,dvipsnames]{color}
\usepackage[normalem]{ulem}

\newcommand{\inE}{\ensuremath{U}}

\begin{document}

\title{Molecular Dynamics in the Multicanonical
  Ensemble:\break{} Equivalence of Wang--Landau Sampling, Statistical
  Temperature Molecular Dynamics, and Metadynamics}

\author{Christoph Junghans}
\author{Danny Perez}
\author{Thomas Vogel}
\email[E-mail: ]{tvogel@lanl.gov}
\affiliation{Theoretical Division T-1, Los Alamos National Laboratory, Los Alamos, NM 87545, USA}

\begin{abstract}
  We show direct formal relationship between the Wang--Landau
  iteration [PRL 86, 2050 (2001)], metadynamics [PNAS 99, 12562
  (2002)] and statistical temperature molecular dynamics (STMD) [PRL
  97, 050601 (2006)], the major work-horses for sampling from
  generalized ensembles. We dem\-onstrate that STMD, itself derived
  from the Wang--Landau method, can be made indistinguishable from
  metadynamics. We also show that Gaussian kernels significantly
  improve the performance of STMD, highlighting the practical benefits
  of this improved formal understanding.
\end{abstract}

\date{April 15, 2014}
\maketitle

\section{Introduction}

Generalized ensemble methods have become the standard techniques to
explore the energy landscape of complex systems~\cite{comment1}.  From
such samplings, the free energy can be obtained, which provides
various thermodynamic insights.  The idea of performing Monte Carlo
(MC) simulations in non-canonical or extended ensembles goes back a
long time. Early milestones include works by Torrie and
Valleau~\cite{torrie1977jcp}, who introduced the so-called Umbrella
Sampling, Challa and Hetherington~\cite{challa88prl,*challa88pra} who
proposed a Gaussian ensemble to interpolate between microcanonical and
canonical views of phase transitions in finite systems, and
Lyubartsev~et~al.~\cite{lyubartsev92jcp} who simulated an expanded
ensemble covering a wide temperature range. Monte Carlo (MC)
simulations in the multicanonial (muca) ensemble, first proposed by
Berg and Neuhaus~\cite{bergneuhaus91plb,*bergneuhaus92prl}, exploit the
umbrella sampling idea by generating an umbrella in a way that a
random walk in energy space is obtained.  Later, Hansmann \textit{et
  al}.~\cite{hansmann96cpl} extended multicanonial MC to molecular
dynamics (MD). Of course, the main technical challenge is the
determination of good umbrellas (multicanonial weights) in order to
achieve a diffusive behavior in energy space. In a related effort,
Wang and Landau (WL) proposed a random-walk
algorithm~\cite{wanglandau01prl,*wanglandau01pre} for MC applications,
in which the density of states, suitable to calculate multicanonical
weights, is estimated on the fly; in fact, over the last decade the WL
method has become the most popular tool for this purpose in the MC
community~\footnote{In a strict sense, the multicanonical ensemble is
  an idealized ensemble. Methods creating extended ensembles which
  eventually converge towards the muca ensemble, such as the original
  muca recursion, WL sampling, or variations of these, are often
  referred to as \textit{flat histogram} methods.  At finite times,
  these ensembles are formally different, however, these differences
  are not critical for our discussion and we will therefore use the
  terms `multicanonical' and `flat-histogram' interchangeably in the
  following.}. Shortly after, Laio and
Parrinello~\cite{laioparrinello02pnas} proposed an MD-based method ---
metadynamics --- to fill up basins of the free-energy surface and
enhance the exploration of configuration space. Using a different
approach, and independently from metadynamics, Kim \textit{et
  al}.~\cite{kim06prl,*kim07jcp} later combined ideas from Hansmann's
multicanonial MD with the WL MC algorithm and put forward a method
known as statistical temperature molecular dynamics (STMD).  Other
combinations of WL and MD using the weighted histogram
method~\cite{nagasima2007pre} and/or smoothing of the estimated
density of states~\cite{shimoyama2011jcp} have been proposed as well.

In this paper we investigate the relationship between WL, STMD and
metadynamics. While all these methods are well established in their
communities, their precise formal relationships has, to the best of
our knowledge, not been thoroughly analysed and, consequently, their
development largely proceeds in parallel. We provide a unified formal
view of these three methods and give the conditions under which they
are equivalent. In particular, we show that STMD and metadynamics
produce, on a timestep per timestep basis, \emph{identical} dynamics
when using consistent initialization and update schemes. This unified
view allows for the transfer of innovation between the different
methods and avoids duplication of efforts in different communities.

\section{Molecular dynamics in the multicanonical ensemble}
\label{sec:muca_ensemble}

In the muca ensemble, one aims at sampling from a \textit{flat}
potential energy distribution $P_{\textrm{muca}}(\inE)$, i.e., one
requires
\begin{equation}
\label{eq:mucaP}
  P_{\textrm{muca}}(\inE)\propto g(\inE)\,w_\textrm{muca}(\inE)=\textrm{const}\,.,
\end{equation}
where $g(\inE)$ is the density of states and $w_\textrm{muca}(\inE)$
are the multicanonical weights, independently of $T$. Obviously, for
this to be realized, the weights have to take the form
\begin{equation}
\label{eq:w_muca}
w_\textrm{muca}(\inE)\propto 1/g(\inE)=\mathrm{e}^{-\ln g(\inE)}=\mathrm{e}^{-k_\textrm{B}^{-1}S(\inE)}\,,
\end{equation}
where $S(\inE)=k_\textrm{B} \ln g(\inE)$ is the microcanonical entropy and
$k_\textrm{B}$ the Boltzmann constant. In the traditional formulation where
only configurational degrees of freedom are taken into account, the
muca weights can be seen as canonical weights at a temperature $T_0$
for an effective potential
\begin{equation}
  \label{eq:Veff}
  V_\textrm{eff}(\inE)=T_0\,S(\inE)\,.
\end{equation}
\vadjust{\break}

\noindent
The interatomic forces for muca MD simulations are obtained from the
gradient of $V_\textrm{eff}(\inE)$:
\begin{align}
f^{\textrm{muca}}_i&=-\frac{\mathrm{d} V_\textrm{eff}(S(\inE(q_1,\ldots,q_{3n})))}{\mathrm{d} q_i}\nonumber\\
&=-T_0\,\frac{\partial S}{\partial \inE}\,\frac{\mathrm{d} \inE(q_1,\ldots,q_{3n})}{\mathrm{d} q_i}\,.
\label{eq:f_muca}
\end{align}
Using the definition of the  microcanonical temperature:
\begin{equation}
\label{eq:microT}
T(\inE)^{-1}=\partial S(\inE)/\partial \inE\,,
\end{equation}
the multicanonical forces become:
\begin{equation}
f^{\textrm{muca}}_i=\frac{T_0}{T(\inE)}\;f_i\,,
\label{eq:fE}
\end{equation}
i.e., muca forces differ from the conventional forces, $f_i$, only by
an energy dependent rescaling factor $\propto 1/T(\inE)$.

Since the multicanonical weights are related to the density of states
(Eq.~\ref{eq:mucaP}), results of a single multicanonical simulation
can be reweighted to obtain canonical averages at any temperature. The
key difficulty in flat-histogram simulations, on the other hand, is to
determine the simulation weights (i.e., the density of states), and
many different approaches have been proposed to address that issue, WL
being one of the most popular. In WL
\cite{wanglandau01prl,*wanglandau01pre}, the density of states
$g(\inE)$ is approximated using a discrete histogram. At each step,
the value of the bin of the instantaneous estimator
$g^\prime(\inE,t)$~\footnote{Primed quantities $y^\prime(x,t)$ will
  generally refer to instantaneous estimators of $y(x)$ in the
  following.}  containing the current energy is updated using a
modification factor $f_{\mathrm{WL}}$ via
\begin{equation}
  \ln\,g^\prime(\inE_\textrm{act},t+\Delta t)=\ln\,g^\prime(\inE_\textrm{act},t)+\ln f_{\mathrm{WL}}\,,
  \label{eq:wl_update}
\end{equation}
where `act' is the actual bin index and $t$ is the MC (or later, MD)
simulation time. Conventionally, $\ln f_{\mathrm{WL}}$ is initially set
to~1 and $\ln\,g^\prime(\inE,t=0)=0$.  Simultaneously, a histogram
$H(\inE)$ of the energy bins visited during the simulation is
accumulated. Once $H(\inE)$ is deemed flat enough, $f_{\mathrm{WL}}$
is decreased, e.g., as $f_{\mathrm{WL}}\to\sqrt{f_{\mathrm{WL}}}$. In
this paper, we are mainly concerned with the first iteration, where
the dynamics are still strongly biased, but it can be shown that, as
$f_{\mathrm{WL}}$ tends to~1, the WL method converges to a~correct
multicanonical
sampling~\cite{wanglandau01prl,*wanglandau01pre,zhou05pre}.

Direct applications of the WL strategy to MD have been
attempted~\cite{nakajima1997jcpb}, however, such efforts were hampered
by numerical stability issues introduced by finite-difference
differentiation of noisy histograms, requiring the introduction of
rather elaborate smoothing procedures~\cite{shimoyama2011jcp}.  To
avoid such complications, Kim \textit{et al}.~\cite{kim06prl,*kim07jcp}
proposed to directly estimate $T(U)$ (cf.~Eqs.~\ref{eq:microT}
and~\ref{eq:fE}) and update $T^\prime(\inE,t)$, which they refer to as
\textit{statistical temperature}, as the MD simulation proceeds
\emph{and} to begin from an initially constant temperature
$T^\prime(\inE,t=0)=T_0>0$ instead of a~constant entropy as done in
WL.  This approach allows for a restriction of the sampled temperature
range, for example to positive values. Except for that key difference,
the STMD scheme is a direct translation of the WL ideas, making muca
MD simulations according to Eq.~(\ref{eq:fE}) feasible.  Applying a
central difference approximation to the derivative in
Eq.~(\ref{eq:microT}), the WL update (Eq.~\ref{eq:wl_update}) then
translates into the following temperature update ($T^\prime(\inE,t)$
is also a binned, discrete function) in the energy bins next to the
currently occupied one:
\begin{equation}
\label{eq:orig_STMD_update}
T^\prime(\inE_{\mathrm{act}\pm1},t+\Delta t)
=\frac{T^\prime(\inE_{\mathrm{act}\pm1},t)}{1\mp\delta_\beta\,T^\prime(\inE_{\mathrm{act}\pm1},t)}\,,
\end{equation}
with $\delta_\beta=k_\mathrm{B} \ln f_{\mathrm{WL}}/2\Delta U$ and
$\Delta U$ being the energy bin width. See
Ref.~\cite{kim06prl,*kim07jcp} for all details.

Various extensions of this single-bin based update scheme are
possible.  For example, one can choose any scalable kernel function
$\gamma\,k(x/\hat{\delta})$ to evolve the entropy estimator
$S^\prime(\inE,t)\propto\ln\,g^\prime(\inE,t)$. The update (which can
now affect an arbitrarily large energy range) then reads:
\begin{equation}
  \label{eq:wl_gauss_update}
  \ln\,g^\prime(\inE,t+\Delta t)=\ln\,g^\prime(\inE,t)+\gamma\,k\left[ (\inE-\inE_\textrm{act})/\hat{\delta}\right].
\end{equation}
This scheme has proven to be particularly useful for Wang--Landau
sampling of joint densities of states, i.e., when performing random
walks in more than one dimension~\cite{zhou06prl}.
The above expression can be cast in terms of an entropy estimator as:
\begin{equation}
  \label{eq:S_sum}
  S^\prime(\inE,t)=\gamma\sum_{t^\ast\leq t} k\left[(\inE-\inE(t^\ast))/\hat{\delta}\right] + S^\prime(\inE,t=0)\,,
\end{equation}
where we use the times $t^\ast$ to index the entropy-update events.
Following STMD, assume the initial guess $S^\prime(\inE,t=0)$ is such
that
\begin{equation}
  \label{eq:S_init}
  \frac{1}{T^\prime(\inE,t=0)}=\frac{\partial S^\prime(\inE,t=0)}{\partial \inE}=\frac{1}{T_0}\,.
\end{equation}
Recalling Eq.~(\ref{eq:f_muca}), we then
get for the muca forces:
\begin{align}
  \nonumber 
  &f^{\textrm{muca}\prime}_i(\inE,t)=T_0\;\frac{\partial
    S^\prime(\inE,t)}{\partial \inE}\,f_i\\ \nonumber
    &\quad=T_0\left(\frac{\partial}{\partial \inE}\,\gamma\sum_{t^\ast\leq t}
    k\left[\frac{\inE-\inE(t^\ast)}{\hat{\delta}}\right] +\frac{\partial S^\prime(\inE,t=0)}{\partial
      \inE}\right)f_i\\
      &\quad=\left(1+\gamma\,T_0\frac{\partial}{\partial \inE}\sum_{t^\ast\leq t}
    k\left[\frac{\inE-\inE(t^\ast)}{\hat{\delta}}\right] \right) f_i \,.
\label{eq:mucaf_kernel_sum}
\end{align}

Taking a step back, we can use this last equation to factorize
$V_\textrm{eff}(\inE)$ (cf. Eq.~\ref{eq:Veff}) into a {\em sum} of the
original potential $\inE$ and a bias potential $V_G$:
$V_\textrm{eff}(\inE)=\inE+V_G$. By inspection (cf. Eqs.~\ref{eq:f_muca}
and~\ref{eq:mucaf_kernel_sum}), we directly get:
\begin{equation}
 V_G^\prime(\inE,t)= \gamma\, T_0\sum_{t^\ast\leq t} k\left[(\inE-\inE(t^\ast))/\hat{\delta}\right]\,,
\end{equation}
i.e., with the proper initial conditions, WL/STMD updates are
equivalent to the construction of an additive bias potential that
takes the form of a simple sum of kernel functions. As we will now
show, this procedure is functionally equivalent to a
metadynamics~\cite{laioparrinello02pnas} approach with the potential
energy as a collective variable. In metadynamics, one also aims at
overcoming free energy barriers, allowing for a random walk in the
collective-variable space~\footnote{$\inE$ is typically not used as the
  collective variable in metadynamics; mainly in cases where one aims
  at estimating the density of states in
  $\inE$~\cite{michelettilaio04prl}. Note that multicanonical and
  other flat-histogram MC methods have also been widely used with
  other collective variables as well, with some examples dating even
  before the introduction of metadynamics. One could mention the bond
  parameter of Potts-like models~\cite{chatelain2005npb}, the Parisi
  overlap parameter for spin glasses~\cite{bergjanke98prl}, or
  interaction parameters in a polymer
  model~\cite{luettmer-strathmann2008jcp} as examples. In general, our
  discussion is independent of the actual choice of this variable, but
  we use $\inE$ for clarity.}.  In order for the system to freely
diffuse with respect to the potential energy, the average
``metadynamics force'' $\phi_{F}$ on the collective variable must
vanish, i.e., the \emph{free energy} landscape $F_{T_0}(\inE)=\inE-T_0
S(\inE)$ must become flat. To that effect, an additive potential
$V_\textrm{G}(\inE)$ is introduced such that
\begin{align}
  \phi_{F}(\inE)=\frac{\partial
    [F_{T_0}(\inE)+V_\textrm{G}(U)]}{\partial \inE} =0\,.
   \label{eq:metadyn_forces}
\end{align}
Clearly, $V_\textrm{G}(\inE)=-F(\inE)$ solves
Eq.~(\ref{eq:metadyn_forces}), which implies
$\inE+V_\textrm{G}(\inE)=T_0 S(\inE)$, up to an arbitrary additive
constant.  Therefore, an energy-based metadynamics simulation simply
reduces to a multicanonical MD simulation in $\inE$
(cf.~Eq.~\ref{eq:Veff}). In practice, metadynamics starts with the
initial guess $V^\prime_\textrm{G}(\inE,t=0)=0$ for the modifying
potential (i.e., also starting the simulation in the canonical
ensemble at temperature $T_0$), which is then gradually updated
following a scheme introduced earlier in the energy landscape paving
method~\cite{hansmann2002prl}.  Typically, Gaussian kernel functions
$k(x/\hat{\delta})\propto
\exp\big[-\frac{1}{2}(x/\hat{\delta})^2\big]$ are used and
$V^\prime_\textrm{G}(\inE,t)$ reads:
\begin{equation}
  \label{eq:Vg}
   V_\textrm{G}^\prime(\inE,t)=w
   \sum_{t^\prime\leq t}\exp\left[-\frac{(\inE-\inE(t^\prime))^2}{2\,\delta \inE^2}\right],
\end{equation}
where $w$ is a tunable constant. The modified interatomic forces are
obtained from the gradient of the modified potential
$\inE(q_1,\ldots,q_{3n})+V_\textrm{G}[\inE(q_1,\ldots,q_{3n}),t]$~\cite{laio08rpp}:
\begin{align}\nonumber
   &f^{\textrm{mod}\prime}_i(U,t)=\frac{\partial \inE}{\partial q_i}
   +\frac{\partial V^\prime_\textrm{G}}{\partial \inE}\,\frac{\partial \inE}{\partial q_i}\\
   &\quad=f_i\left(1+\frac{\partial}{\partial \inE}\,w
     \sum_{t^\prime\leq t}\exp\left[-\frac{(\inE-\inE(t^\prime))^2}{2\,\delta \inE^2}\right]
   \right),
   \label{eq:metadyn_particle_forces}
\end{align}
which is indeed identical to Eq.~(\ref{eq:mucaf_kernel_sum}) when
$\gamma\,k(x/\hat{\delta})$ is a Gaussian kernel function with
$w=\gamma\,T_0$ and when using the same time sampling points
$t^\prime$ and $t^\star$, respectively.

\section{Results and Discussion}

During the last decade, there have been multiple independent
algorithmic advances in the MC and MD communities that lead to
significant improvement of the major generalized ensemble methods, see
Refs.~\cite{mitsutake01ps,singh12cbe} for some examples, and the
introduction of STMD~\cite{kim06prl,*kim07jcp} was a major step in
bridging the gap between MC and MD. Our demonstration that STMD and
metadynamics can be made identical should further facilitate
technological transfers between both communities.

\begin{table}[t]
  \centering
  \caption{Average times to complete first iteration, i.e., create a flat
    histogram in a given energy range. Statistical errors were estimated through multiple independent runs.}
\begin{tabular*}{\columnwidth}{@{\extracolsep{\fill}}lr}\hline
  Method & time in ns \\ \hline
  original STMD ($\Delta U=2$\,eV) & $81.3\pm 27.6$\\
  Gaussian kernel, $\hat{\delta}=\Delta U/\sqrt{2}$ & $88.7\pm 36.6$\\
  Gaussian kernel, $\hat{\delta}=2\,\Delta U/\sqrt{2}$ & $38.8\pm 18.7$\\
  Gaussian kernel, $\hat{\delta}=3\,\Delta U/\sqrt{2}$ & $25.9\pm 23.3$\\ \hline
\end{tabular*}
\label{tab:1}
\end{table}
\begin{figure*}[t]
  \includegraphics[width=.8\textwidth]{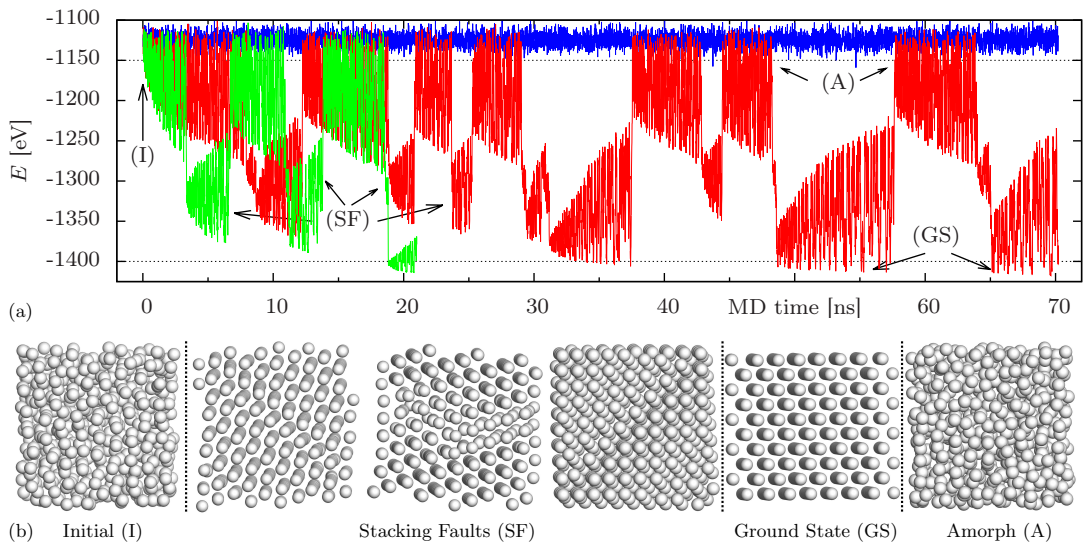}
  \caption{(a) Time series of the first iteration of original (red)
    and Gaussian-kernel STMD runs (green), compared to canonical MD at
    $T=T_0$ (blue). (b) Snapshots of actual atomic configurations
    sampled during the runs. }
\label{fig:res1}
\end{figure*}
The use of Gaussian kernels, as done in metadynamics, in STMD is the
most obvious example of such a~transfer.  For illustrative purposes
(see the Appendix for another example), potential gains are
demonstrated using a system consisting of $500$ silver atoms at
constant particle density $\rho=0.0585$\,\AA$^{-3}$, interacting via
an em\-bedded-atom potential~\cite{williams06msmse}.  We use the
stochastic Velocity-Verlet algorithm~\cite{melchionna07jcp} with a
time step of 2\,fs and a~Langevin thermostat at $T_0=3500\,\text{K}$
and apply periodic boundary conditions. We use the original STMD
method, where the statistical temperature is updated according to a
single-bin update of the entropy (via Eq.~\ref{eq:orig_STMD_update}),
and compare with the Gaussian-kernel version where we directly solve
Eq.~(\ref{eq:microT}). Applying Eq.~(\ref{eq:wl_gauss_update}), this
leads to the following temperature update:
\begin{align}\nonumber
  &T^{-1\prime}(\inE,t+\Delta t)=\frac{\partial
    S^\prime(\inE,t+\Delta t)}{\partial \inE}
  \\ \nonumber &=k_\textrm{B}\,
  \frac{\partial}{\partial \inE}\left[\ln g^\prime(\inE,t)+\gamma\,\textrm{e}^{-(U-U(t))^2/2\hat{\delta}^2}\right]\\
  &=T^{-1\prime}(\inE,t)-2\gamma
  k_\mathrm{B}\left[(\inE-\inE(t))/2\hat{\delta}^2\right]\textrm{e}^{-(U-U(t))^2/2\hat{\delta}^2}\,.
 \label{eq:Gauss_STMD_update}
\end{align}
We apply Gaussian kernels of different widths, which we measure in
units of the energy bin width $\Delta \inE$ used in the original-STMD
run. That is, for $\hat{\delta}=n\Delta U/\sqrt{2}$ the kernel
function drops to $\gamma/e$ at the centers of the $n$th nearest
neighbor energy bins. $\gamma$ takes the role of $\ln f_{\mathrm{WL}}$
(cp. Eq.~\ref{eq:wl_update}) and can be chosen much smaller than for
WL simulations~\cite{kim06prl,*kim07jcp}, we initially set
$\gamma=3.5\times10^{-3}$. We furthermore use a cutoff of $10\Delta
\inE$ at both sides of the Gaussian in all cases, but verified that
the actual choice of the cutoff does not systematically affect the
results (see Appendix for a more detailed discussion and data). The
energy-histogram bin width is identical in all cases and the energy
histogram itself is always updated by increasing single bins, i.e.,
the Gaussian kernels are not applied for recording the histogram of
visited energies. Also, the flatness criterion is identical for all
runs.
In Table~\ref{tab:1} we show the average times needed for different
runs to fulfill the histogram flatness criterion, i.e., to finish the
first WL iteration and, in particular, to visit all energy levels.
The result clearly shows that the width of the Gaussian kernel
influences how fast the system is driven through energy space, and
that wider kernels provide a significant speedup. In
Fig.~\ref{fig:res1}\,(a), we show time series for the first iteration
from two runs, applying the original STMD and a Gaussian kernel run
with width $3\Delta \inE/\sqrt{2}$, respectively.  For the latter
case, the system moves from the initial~(I), amorphous configuration
via low-energy crystalline states exhibiting stacking faults (SF) to
the perfectly ordered ground state (GS; see Fig.~\ref{fig:res1}\,b for
visualizations) in just about 20\,ns.  Concomitantly, extensive
thermodynamic information is gathered. Also note that the use of
continuous kernel functions, rather than updates of binned estimators,
allows, in principle, for an arbitrarily fine-meshed estimation of
$T(U)$ without systematically influencing the algorithmic runtime.

Many other improvements can be considered and parallel efforts in the
different communities are a common occurrence.  For example, it has
been shown that the WL energy probability distribution is attracted to
the vicinity of the uniform distribution, i.e., that the algorithm
converges to the right solution~\cite{zhou05pre}. By introducing a
height-reduction scheme for the Gaussian kernels~\cite{barducci08prl}
similar statements should be available for MD methods. A more recent
development in the metadynamics community concerns adaptive
Gaussians~\cite{branduardi2012jctc}, where the form of the update to
the bias potential depends on local properties of the underlying
free-energy surface. Similar ideas of applying different entropy
updates in Wang--Landau simulations have circulated~\cite{yin12cpc}
and an ad-hoc method for nonuniform binning of energy levels has been
recently and independently implemented~\cite{koh13pre}. To mention a
final example, in efforts to develop massively parallel
implementations, multiple parallel walkers have been
\emph{simultaneously} deployed to update a bias potential in
metadynamics~\cite{raiteri06jpcb}. However, systematic errors,
unnoticed in Ref.~\cite{raiteri06jpcb}, were detected when exactly
the same approach was independently applied in the MC
community~\cite{yin12cpc}.  Joining insights from both studies might
lead to further improvements. In fact, a generic parallel scheme based
on replica exchanges, which avoids such artificial bias, was recently
introduced and applied in both\break
communities~\cite{kim09jcp,kim12jpcb,vogel13prl,gaivogel13jcp}.

\section{Summary}

We aim at consolidating the developments in the different areas of
generalized ensemble MC and MD sampling by demonstrating that three
popular methods, namely Wang--Landau, Statistical Temperature
Molecular Dynamics, and Metadynamics, are formally equivalent upon a
consistent choice of initial conditions and update rules.
Specifically, we have shown that STMD, a translation of the
Wang--Landau method into the MD language, augmented by the
introduction of kernel updates of the statistical temperature becomes
completely identical to metadynamics, i.e., both methods give
identical dynamics on a timestep by timestep basis.  The focus of this
paper is on this explicit equivalence; discussions concerning the
overall convergence behavior and analogies in that regard between
different strategies in Wang--Landau sampling and metadynamics can be
found in the literature, see
Refs.~\cite{michelettilaio04prl,zhou05pre,belardinelli2008pre,crespo2010pre}.
We believe that a consistent view of flat-histogram methods as
presented here is beneficial to foster transfer of ideas between the
respective \hbox{communities}.\vskip-.3\baselineskip
\begin{figure*}[t]
  \includegraphics[width=.9\textwidth]{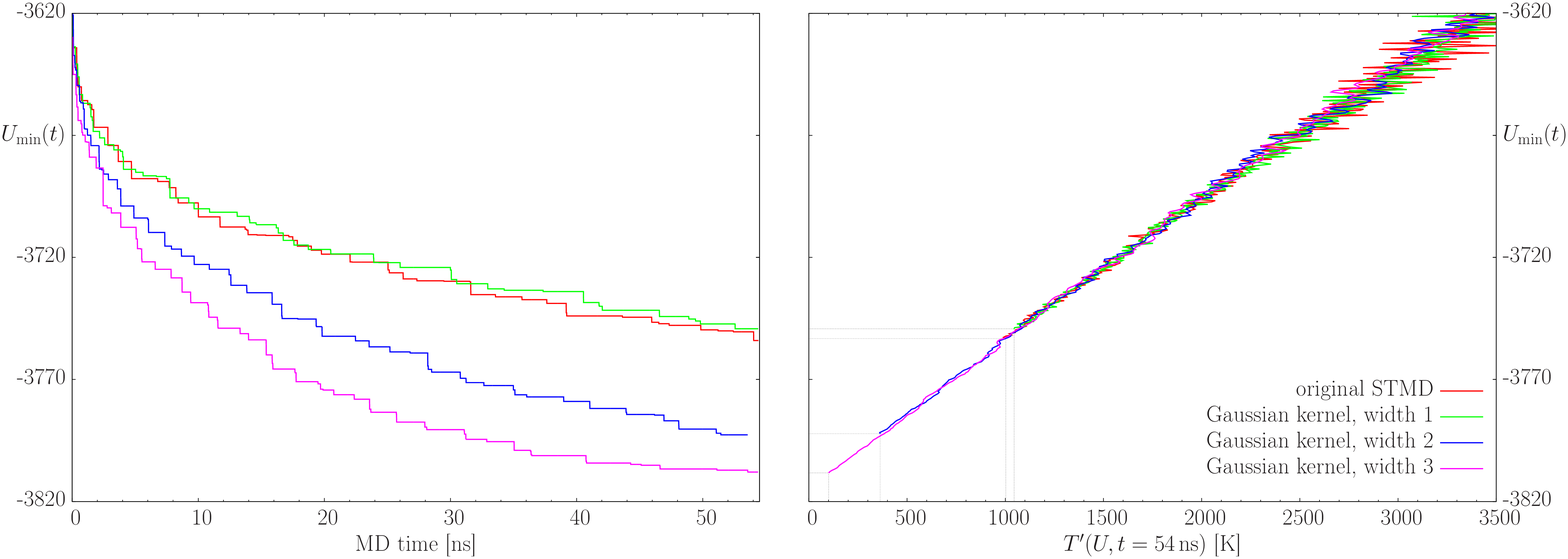}\\
  \hskip-11cm\vskip-6.1cm
  \includegraphics[width=.18\textwidth]{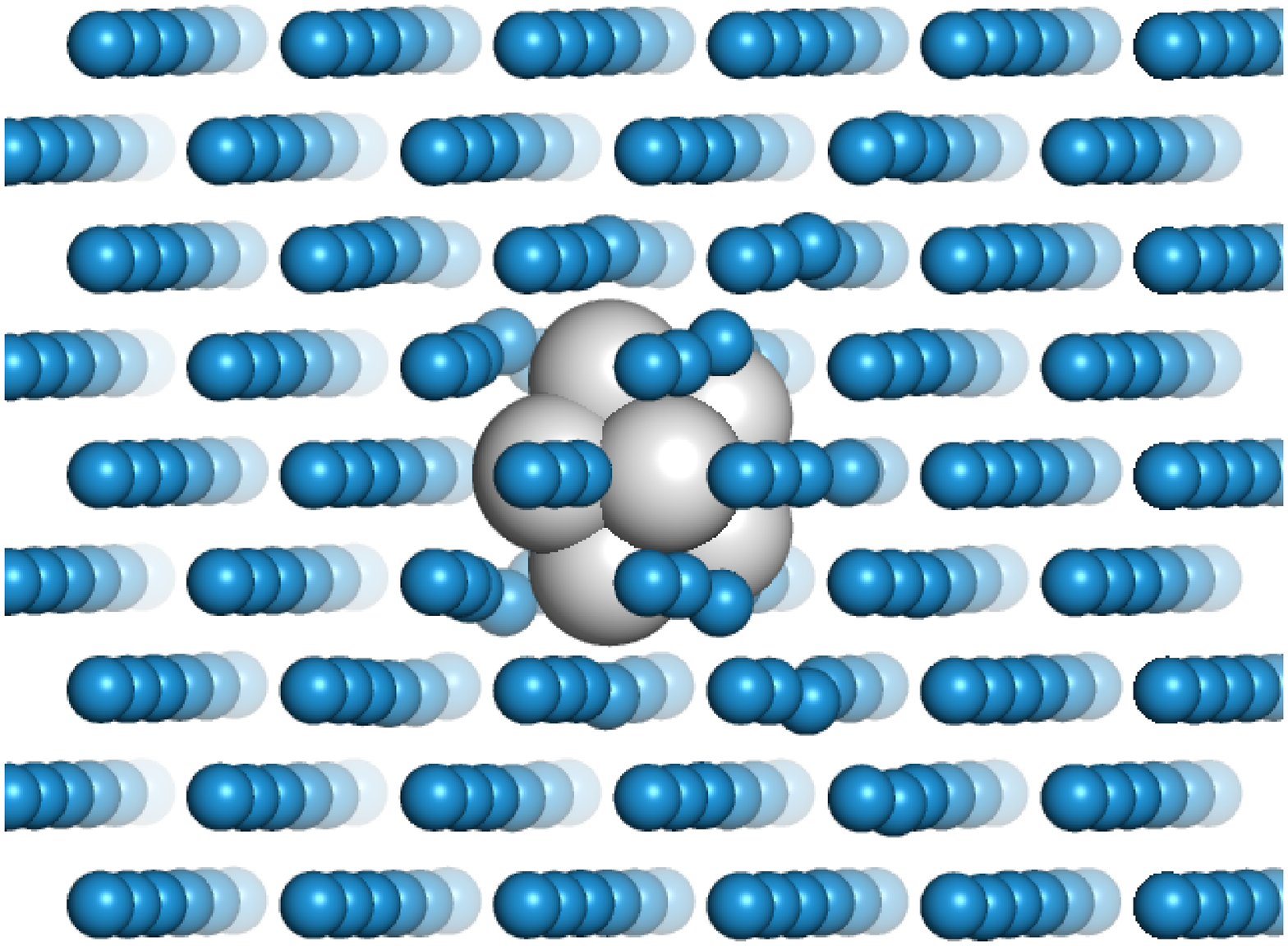}\\
  \vskip3cm
  \caption{(Left) Minimal energy values expored by STMD as a function of
    simulation time for different kernel functions. (Right) The
    corresponding statistical temperatures after $t\approx54\,$ns.
    Dashed lines mark minimal temperatures corresponding to the
    minimal energies explored at that time. (Inset) illustration of
    the physical system in its ground state.}
\label{fig:A1}
\end{figure*}
\begin{figure*}[t]
  \includegraphics[width=.9\textwidth]{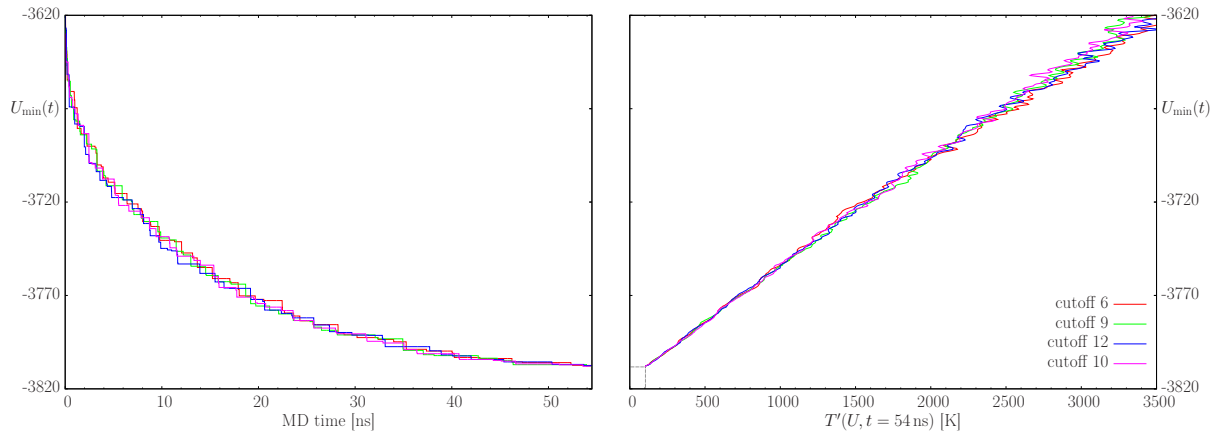}
  \caption{(Left) Minimal energy values expored by STMD depending on
    simulation time for different cutoffs of Gaussian kernel
    functions. (Right) The corresponding statistical temperatures
    after $t\approx54\,$ns.}
\label{fig:A2}
\end{figure*}

\begin{acknowledgments}\noindent
  We thank A.F. Voter for discussions during the entire project
  and Y.W. Li, L. Vernon, J. Kim, J.E. Straub, and T. Keyes for
  critical reading of the manuscript. TV and DP acknowledge funding by
  Los Alamos National Laboratory's (LANL) Laboratory Directed Research
  and Development ER program, and CJ by a LANL Director's fellowship.
  Assigned: LA-UR 13-29519.  LANL is operated by Los Alamos National
  Security, LLC, for the National Nuclear Security Administration of
  the U.S. DOE under Contract DE-AC52-06NA25396.
\end{acknowledgments}

\section*{Appendix}

Supporting our statements made above, we present here additional data
showing the effect of the width of Gaussian kernel functions on the
performance of STMD and how a~cutoff of those kernels, typically used
in practice, affects the results.

\vspace*{-.5\baselineskip}
\subsection*{Further examples of the speedup due to Gaussian kernel functions}
\vspace*{-.5\baselineskip}

\noindent
We show in Fig.~\ref{fig:A1} performance results of STMD simulations
on a system composed of a tungsten bcc crystal containing six helium atoms.
The clustering of He in W presents an important technological
challenge because it can lead to severe microstructural modifications
in expected operating conditions of future magnetic-confinement fusion
reactors. Indeed, as interstitial He cluster grow, they reach a point
where they are able to eject W atoms from the lattice and condense
into the resulting vacancies, creating the nuclei of a bubble that can
then grow and disrupt the structure of the material
(see Refs.~\cite{Perez2014,Vogel2014} for further details and 
results on that system). The reference temperature is set to $T_0=3500\,$K (upper
temperature boundary of right plot in Fig.~\ref{fig:A1}), which
corresponds to a canonical mean energie of $\langle
U\rangle_\mathrm{canonical}\approx-3620\,$eV (upper energy boundary in plots
in Fig.~\ref{fig:A1}). When applying wider Gaussian kernel functions
(within reasonable limits), the energy and, hence, statistical
temperature range is explored much faster, which confirms our claim in
the main article. While the original STMD method explores temperatures
down to $T\approx1000\,$K, temperatures down to $T\approx100\,$K are
visited when applying Gaussian kernels of width $\hat{\delta}=3\Delta
U/\sqrt{2}$.

\subsection*{Influence of the cutoff of Gaussian kernel functions}

\noindent
One can reasonably expect that the actual choice of the cutoff of the
Gaussian kernel should not affect physical results. Indeed, the
original STMD can be seen as the limit of a short cutoff, and, as
shown in Fig.~\ref{fig:A1} (right panel), its statistical temperature
agrees with that of longer cutoff kernels. This is further
demonstrated in Fig.~\ref{fig:A2}, where the results presented above
were reproduced with different values of the cutoff for Gaussian
kernel functions of width $\hat{\delta}=3\Delta U/\sqrt{2}$ (lowest
curve in Fig.~\ref{fig:A1}). Results coincide (right panel) and we
observe no notable, systematic difference in the performance in terms
of the explored energy and temperature ranges at all times during the
runs (left panel). Lowest temperatures explored after $t\approx54\,$ns
are all in the range $104\leq T \leq 107\,$K. One should however note
that using a long cutoff could affect the very low temperatures, as
the statistical temperature can approach zero with a finite slope.
This discontinuity would be somewhat smoothed out by Gaussian kernels
with a large cutoff.

\end{document}